\documentclass[aps,twocolumn,prl,floatfix,longbibliography,superscriptaddress,amssymb,amsmath]{revtex4-1}
\pdfoutput=1
\usepackage{graphicx,braket,float}
\usepackage{amsmath}
\usepackage{amsfonts}
\usepackage{amsbsy}
\usepackage{xcolor}
\usepackage{soul}
\usepackage{bm}
\usepackage[normalem]{ulem}
\usepackage[unicode]{hyperref}
\hypersetup{
    unicode=true, % non-Latin characters in Acrobat?s bookmarks
    plainpages=false,
    colorlinks=true,% false: boxed links; true: colored links
    linkcolor=blue,% color of internal links
    citecolor=blue,% color of links to bibliography
    filecolor=blue,% color of file links
    urlcolor=blue% color of external links
}
\newcommand{\cbEL}{\boldsymbol{\mathbf{\cal E}}}

\begin{document}

\title{A superatom picture of collective nonclassical light emission and dipole blockade in atom arrays}
\author{L. A. Williamson}
\affiliation{Department of Physics, Lancaster University, Lancaster LA1 4YB, United Kingdom}
\author{M. O. Borgh}
\affiliation{Faculty of Science, University of East Anglia, Norwich NR4 7TJ, United Kingdom}
\author{J. Ruostekoski}
\affiliation{Department of Physics, Lancaster University, Lancaster LA1 4YB, United Kingdom}
\date{\today}

\begin{abstract}
We show that two-time, second-order correlations of scattered photons from planar arrays and chains of atoms display nonclassical features that can be described by a superatom picture of
the canonical single-atom $g_2(\tau)$ resonance fluorescence result. For the superatom, the single-atom linewidth is replaced by the linewidth of the underlying collective low light-intensity eigenmode. 
Strong light-induced dipole-dipole interactions lead to a correlated response, suppressed joint photon detection events, and dipole blockade that inhibits multiple excitations of the collective atomic state.
For targeted subradiant modes, nonclassical nature of emitted light can be dramatically enhanced even compared with that of a single atom.
\end{abstract}

\maketitle

The first direct evidence for the quantum nature of light was observed in resonance fluorescence of an atom~\cite{Carmichael_1976,kimble1977,kimble1978,dagenais1978,walls1979}, defining a significant historical milestone in quantum optics.
Such quantum correlations can be identified by measuring the second-order correlation function for the emitted field that represents a joint probability of two photon detection events appearing a time $\tau$ apart and can be defined as
\begin{equation} \label{eq:g2}
g_2(\tau)\equiv\lim_{t\rightarrow\infty}\frac{\langle\,:\!\hat{n}(t+\tau)\hat{n}(t)\!:\,\rangle}{\langle\hat{n}(t)\rangle^2},
\end{equation}
where $:\,:$ denotes normal ordering and $\hat{n}(t)$ is the number operator for detected photons. Classically, $g_2(0)\ge g_2(\tau)$; hence $g_2(0)<g_2(\tau)$ implies quantum correlations in the photon emission, and also defines antibunched photon emission~\cite{mandel1990,paul1982}. 

Going beyond a single atom, in a noninteracting ensemble atoms will emit photons independently, leading to an adulteration of the single-atom photon antibunching that (neglecting interferences) scales as $1-N^{-1}$, with  the atom number $N$~\cite{kimble1978,jakeman1977,carmichael1978}. 
Correlated excitations for atomic ensembles have been observed for highly-excited Rydberg atoms in the microwave regime. The correlated response is generated by dipolar interactions that inhibit
transitions into all but singly-excited states, representing dipole blockade~\cite{Jaksch00,Lukin01,Urban09,Grangier09,Saffman10,Schauss12,ripka2018}, with applications to scalable quantum logic gates.

In dense ensembles of cold atoms, also light-mediated interactions between the atoms can lead to drastic and unexpected phenomena~\cite{Javanainen2014a,Skipetrov14,guerin16_review,Javanainen17} as multiple resonant scattering events give rise to a correlated response.  Correlations can emerge even for the classical optical regime in the limit of low light intensity (LLI) of an incident laser~\cite{Morice1995a,Ruostekoski1997a}, and the quest for observing the effects of strong light-mediated interactions is attracting considerable attention ~\cite{BalikEtAl2013,CHA14,Pellegrino2014a,wilkowski,Jennewein_trans,Ye2016,Jenkins_thermshift,vdStraten16,Guerin_subr16,Machluf2018,Dalibard_slab,Bettles18}. Regular arrays of atoms are particularly interesting for the exploration and manipulation of collective optical responses, as more recently studied also in the quantum regime~\cite{Ritsch_subr, Olmos16,Zhang2018, Grankin18,Guimond2019, bettles2019,ballantine2019, Needham19, Qu19,williamson2020, Zhang20,Henriet2018, Asenjo-Garcia19}. Transmission-resonance narrowing due to collective subradiance in the classical limit in a planar optical lattice was already observed~\cite{rui2020} and other related experiments are rapidly emerging~\cite{Glicenstein20}. 

Here we show that photon emission events from planar arrays and chains of atoms can still be described by the single isolated atom picture, representing a collective response of the entire atomic ensemble as one \emph{superatom}.
By resonantly targeting LLI collective excitation eigenmodes, we show that even at high light intensities the many-atom joint photon emission $g_2(\tau)$ displays the same functional form as the single isolated atom $g_2(\tau)$ of Eq.~\eqref{eq:g2}, but with the single atom linewidth replaced by the linewidth of the targeted LLI collective mode. 
We find that for sufficiently small lattice spacings strong light-induced interactions can increase antibunching by establishing correlations between the atoms that represent inhibited multiple excitations of the collective state of the atoms, or \emph{dipole blockade}.
Remarkably, for underlying LLI eigenmodes for which the resonance linewidth is much narrower than the one for an isolated atom (subradiance), the nonclassical nature of emitted light can be dramatically enhanced to much longer time scales even compared with those of a single atom.

We consider two-level atoms with the dipole matrix element $\mathbf{d}$, coupled by light-mediated interactions and subject to an incident laser field. The atom dynamics in the rotating-wave approximation follows from the many-body quantum master equation (QME) for the reduced density matrix~\cite{Lehmberg1970,agarwal1970,supplement},
\begin{equation}\label{masterEq}
  \begin{split}
    \frac{d\rho}{dt}= &-\frac{i}{\hbar}\sum_j[H_j,\rho]+i\sum_{j\ell(\ell\ne j)}\Delta_{j\ell}[\hat{\sigma}_j^+\hat{\sigma}_\ell^-,\rho]\\
    &+\sum_{j\ell}\gamma_{j\ell}\left(2\sigma_j^-\rho\sigma_\ell^+-\sigma_\ell^+\sigma_j^-\rho-\rho\sigma_\ell^+\sigma_j^-\right)
  \end{split}
\end{equation}
with the atomic operators $\hat{\sigma}_j^+= (\hat{\sigma}_j^-)^\dagger =  |e\rangle_{j}\mbox{}_{j}\langle g|$, $\hat{\sigma}_j^{ee}=\hat{\sigma}_j^+\hat{\sigma}_j^-$, for ground $|g\rangle_{j}$ and excited $|e\rangle_{j}$ states of atom $j$ located at $\mathbf{r}_j$ and
\begin{equation}
H_j\equiv -\hbar\delta\hat{\sigma}_j^{ee}-\mathbf{d}\cdot\cbEL^+(\mathbf{r}_j)\hat{\sigma}_j^+-\mathbf{d}^*\cdot\cbEL^-(\mathbf{r}_j)\hat{\sigma}_j^-.
\end{equation}
We take the positive-frequency component $\cbEL^+(\mathbf{r})=\frac{1}{2}\mathcal{E}_0 e^{i\mathbf{k}\cdot\mathbf{r}}\hat{\mathbf{e}}$ of the laser field to be a monochromatic plane wave of frequency $\omega=kc=2\pi c/\lambda$ and wavevector $\mathbf{k}$, detuned from the single-atom transition frequency $\omega_0$ by $\delta\equiv\omega-\omega_0$.  The light and atomic field amplitudes are here defined as slowly varying with the rapid oscillations at the laser frequency factored out. The light-mediated interactions between the atoms have both coherent $\Delta_{j\ell}$ and dissipative $\gamma_{j\ell}$ contributions [$\gamma_{jj}=\gamma \equiv |\mathbf{d}|^2k^3/(6\pi\hbar\epsilon_0)$ is the single atom linewidth]. These are the real and imaginary parts, respectively, of $\mathbf{d}^*\cdot\mathsf{G}(\mathbf{r}_j-\mathbf{r}_\ell)\mathbf{d}/\hbar\epsilon_0$, with $\mathsf{G}(\mathbf{r})$ the dipole radiation kernel of a point dipole at the origin~\cite{Jackson,supplement}.

In the limit of LLI the dynamics
reduces to that of classical coupled dipoles~\cite{Javanainen1999a,Lee16}. 
In this regime we may describe~\cite{supplement} the optical response using LLI collective radiative excitation eigenmodes $u_m$ of $\mathcal{H}_{j\ell}=\Delta_{j\ell}+i\gamma_{j\ell}$ (with $\Delta_{jj}\equiv 0$), with the complex eigenvalues $\zeta_m+i\upsilon_m$ representing the collective linewidth $\upsilon_m$ and line shift $\zeta_m$ from the single-atom resonance. The linewidths can span many orders of magnitude, from extremely subradiant to superradiant~\cite{Jenkins2012a,Jenkins_thermshift,Sutherland1D}. 

To calculate the rate of the detected photons for the second-order correlation function $g_2(\tau)$ of Eq.~\eqref{eq:g2} we assume all the scattered photons are detected 
and integrate $\hat{n}(t)=(2\epsilon_0 c/\hbar\omega_0)\int_S dS\,\hat{\mathbf{E}}_\text{sc}^-(\mathbf{r},t)\cdot\hat{\mathbf{E}}_\text{sc}^+(\mathbf{r},t)$ over a closed surface enclosing the atoms to give $\hat{n}=2\sum_{j\ell}\gamma_{j\ell}\sigma_j^+\sigma_\ell^-$~\cite{supplement}, 
where $\epsilon_0\hat{\mathbf{E}}_\text{sc}^+(\mathbf{r},t)=\sum_j \mathsf{G}(\mathbf{r}-\mathbf{r}_j)\mathbf{d}\hat{\sigma}_j^-(t)$ denotes the scattered electric field summed over all the atoms. 
For a single isolated atom, a closed expression for $g_2(\tau)$ can be derived analytically and is given by~\cite{Carmichael_1976,carmichael1976},
\begin{equation}\label{eq:g2single-atom}
g_2^{(\gamma,\kappa)}(\tau)\equiv 1-e^{-3\gamma \tau/2}\left(\cosh \kappa \gamma \tau+\frac{3}{2}\frac{\sinh \kappa \gamma \tau}{\kappa}\right),
\end{equation}
where $\kappa \equiv \frac{1}{2}\sqrt{1-8I_\text{in}/I_s}$, and $I_\text{in}\equiv \epsilon_0 c|\mathcal{E}_0\hat{\mathbf{e}}\cdot\hat{\mathbf{d}}|^2/2$ and $I_s\equiv\hbar ck^3\gamma/6\pi$ are the incident light and saturation intensities, respectively. For $g_2^{(\gamma,\kappa)}(0)=0$ and $\lim_{\tau\rightarrow\infty}g_2^{(\gamma,\kappa)}(\tau)=1$; a single isolated atom therefore shows photon antibunching, a manifestation of the fact that an atomic energy level can contain at most a single excitation.

For the many-body system, $g_2(\tau)$ [Eq.~\eqref{eq:g2}] in general needs to be evaluated by first solving the QME \eqref{masterEq} numerically. The existence of nonclassical effects
for a many-atom ensemble is less obvious than in the single-atom case. This can be illustrated by a simple counting example of $N$ independently emitting, noninteracting atoms: Neglecting interferences then yields $g_2(\tau)=1+N^{-1}[g_2^{(\gamma,\kappa)}(\tau)-1]$, indicating a rapidly reduced photon antibunching as a function of the atom number, as photons from independently emitting atoms wash out the correlations.

For the case of strong cooperative coupling of closely-spaced atoms we have a strongly correlated quantum many-body system with long-range dipole-dipole interactions. 
While we have also numerically calculated $g_2$ for such situations, our key observation is that for several strongly correlated regimes of interest, Eq.~\eqref{eq:g2single-atom} remarkably can still provide a qualitative description for emitted photon correlations that also exhibit nonclassical scattered light and inhibited multiple excitations (dipole blockade) even for increasing atom numbers. This is because atoms collectively respond as one giant \emph{superatom}, where effectively the single-particle resonance linewidth is replaced by the resonance linewidth of the dominant underlying LLI collective excitation eigenmode.

The dominant eigenmode in a regular array is determined by the resonance frequency and phase-matching profile with the incident field. We find then that the many-body $g_2(\tau)$ obeys a functional form analogous to Eq.~\eqref{eq:g2single-atom},
\begin{equation}\label{eq:superatom}
g_2(\tau)\approx 1+b\left[g_2^{(\upsilon,\kappa^\prime)}(\tau)-1\right],
\end{equation}
where $\upsilon=\upsilon_\ell$ is the linewidth of the resonant LLI eigenmode $u_\ell$ (found by diagonalising $\mathcal{H}_{j\ell}$~\cite{supplement}) and $\kappa^\prime \equiv \frac{1}{2}\sqrt{1-8\mathcal{I} I_\text{in}/I_s^\prime}$, with $I_s^\prime\equiv\hbar ck^3\upsilon/6\pi$.
The overlap of the drive field with $u_\ell$, $\mathcal{I}=|\sum_j e^{-i\mathbf{k}\cdot\mathbf{r}_j}u_\ell (\mathbf{r}_j)|^2$, represents the sum of the coupling strengths of light over all the atoms  and can for uniform targeted modes with perfect phase-matching be replaced by $N$, reflecting the collective $N$-enhancement of the response. There is an overall normalization in Eq.~\eqref{eq:superatom} by $b\approx 1-g_2(0)$ that accounts for nonclassical light emission at zero delay due to many-body correlations. When $b>N^{-1}$, these are enhanced compared to the noninteracting, noninterfering case.

\begin{figure}
\includegraphics[trim=0.7cm 10.4cm 0cm 11.5cm,clip=true,width=0.5\textwidth]{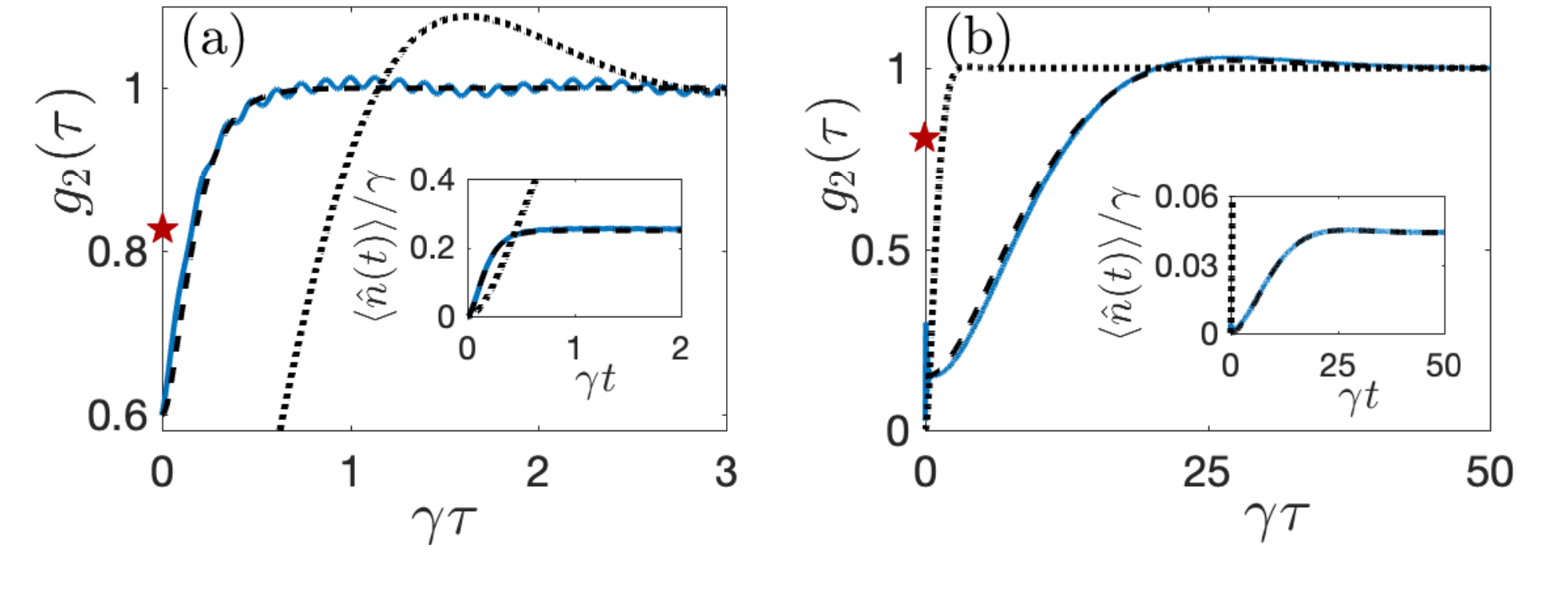}\vspace{-0.6cm}
\caption{\label{2d}Superatom picture and nonclassical light scattering for a $3\times 3$ atom array (lattice spacing $a=0.1\lambda$) with a drive field resonant with (a) the uniform superradiant ($\upsilon\approx 7.6\gamma$, $NI_\text{in}=2I_s$) and (b) a subradiant ($\upsilon\approx 0.091\gamma$, $NI_\text{in}=0.5I_s$) LLI collective eigenmode; $g_2(\tau)$ for the full quantum solution (blue solid line), superatom (black dashed line), and single isolated atom (black dotted line). The red star marks the noninteracting, interfering result of $g_2(0)$, showing that interactions substantially enhance photon antibunching. For subradiant mode the nonclassical emission is enhanced compared with a single atom.
Insets show the corresponding photon detection rates.}
\end{figure}

In the numerics, we consider 2D square arrays of atoms in the $xy$ plane and 1D chains along the $x$ axis, with the incident light direction $\hat{\mathbf{k}}=\hat{\mathbf{z}}$, polarized along the atomic dipoles $\hat{\mathbf{d}}=\hat{\mathbf{x}}$. 
We solve the QME by directly integrating Eq.~\eqref{masterEq} or
by unraveling the evolution into stochastic quantum trajectories of state vectors~\cite{dalibard1992,Tian92,Dum92,carmichael1993,supplement}.  

We demonstrate nonclassically scattered light from a strongly interacting $3\times 3$ planar array of atoms  in the two-time correlation function in Fig.~\ref{2d}, where the nonclassicality of the photon emission is strongly \emph{enhanced} due to interactions. 
This corresponds to inhibited multiple excitations of the collective atomic state due to light-mediated dipole-dipole interactions, representing \emph{dipole blockade} of optical transitions, analogous to collective suppression of microwave Rydberg excitations~\cite{Lukin01}. 
The drive, which is uniform across the plane, couples most strongly to the most superradiant LLI eigenmode with no phase variation across the atoms. We show that the superatom picture (SAP) [Eq.~\eqref{eq:superatom}] provides an excellent description of $g_2(\tau)$ for light resonant with this mode ($\upsilon\approx 7.6\gamma$, $\mathcal{I}\approx 0.98 N$) [Fig.~\ref{2d}(a)]. The antibunching delay time is much shorter than that of a single atom.

The incident light can also be tuned to target a subradiant eigenmode. Here we consider the eigenmode with the fourth broadest resonance, with $\upsilon\approx 0.091\gamma$ and $\mathcal{I}\approx 0.015 N$. We find that the SAP again accurately describes the dynamics [Fig.~\ref{2d}(b)]. The mode is approximately $u_\ell(\mathbf{r}_j)\approx 1.4\cos(\pi \hat{\mathbf{x}}\cdot\mathbf{r}_j/2a)-0.12$ with the constant giving rise to nonorthogonality of the eigenmodes. The linewidths of the superradiant and subradiant eigenmodes differ by two orders of magnitude, resulting in very different responses, and in both cases radically departing from the single-atom result. The substantially larger values of $1-g_2(0)$ compared to those of noninteracting atoms show enhanced antibunching due to interactions. In the subradiant case nonclassical effects are \emph{enhanced} compared even with those of a single atom, with the nonclassical delay time of $g_2$ approximately 10 times larger than that of a single atom. Subradiant excitations can therefore provide much extended antibunching time scales compared with Rydberg atom based vapor cell devices~\cite{ripka2018}, also avoiding two-photon excitations and the involvement of highly excited Rydberg states that are sensitive to electric and magnetic field gradients.

The SAP also provides an excellent description of the transient photon scattering rate $\langle\hat{n}(t)\rangle$ (insets to Fig.~\ref{2d} and Fig.~S1 in~\cite{supplement}). The SAP for the photon scattering rate is $\langle\hat{n}(t)\rangle\approx n^{(\upsilon,\kappa^\prime)}(t)$, where $n^{(\gamma,\kappa)}(t)\equiv [I_\text{in}/(I_\text{in}+I_s)]g_2^{(\gamma,\kappa)}(t)$ is the photon scattering rate for a single, isolated atom~\cite{mollow1969b,carmichael1976}. 

\begin{figure}
\includegraphics[width=0.98\columnwidth]{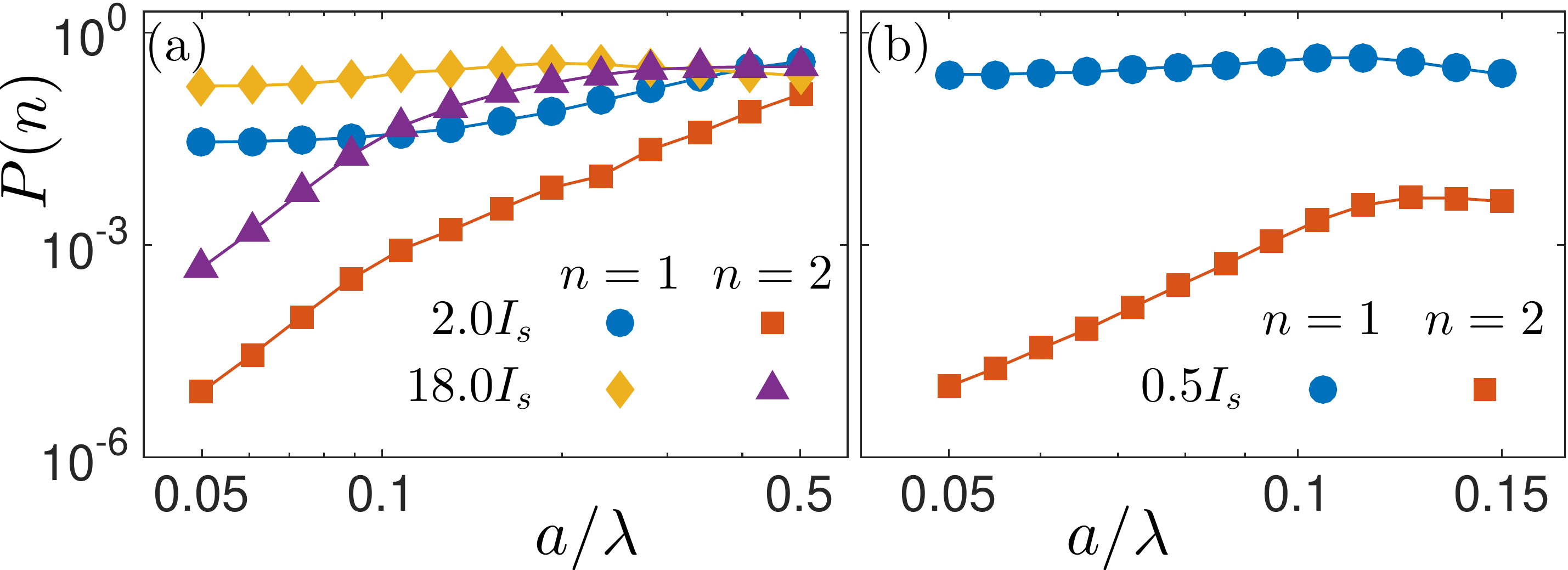}
%\vspace{-0.7cm}
\caption{\label{blockade} Dipole blockade by the occupation weights $P(n)$ of states with $n=1$ and 2 excited atoms as a function of the lattice spacing $a$ in a $2 \times 3$ array, with (a) the superradiant, (b) subradiant LLI eigenmode targeted. For small $a$, $n=2$ states are suppressed by the blockade, regardless of intensity, but the blockade is weakened for larger $a$ and the occupation increases dramatically while the weight of $n=1$ states changes little in comparison. Drive intensity $NI_\text{in}$ given as a multiple of $I_s$.}
\end{figure}

\begin{figure}
\includegraphics[trim=1cm 10cm 0cm 10.6cm,clip=true,width=0.5\textwidth]{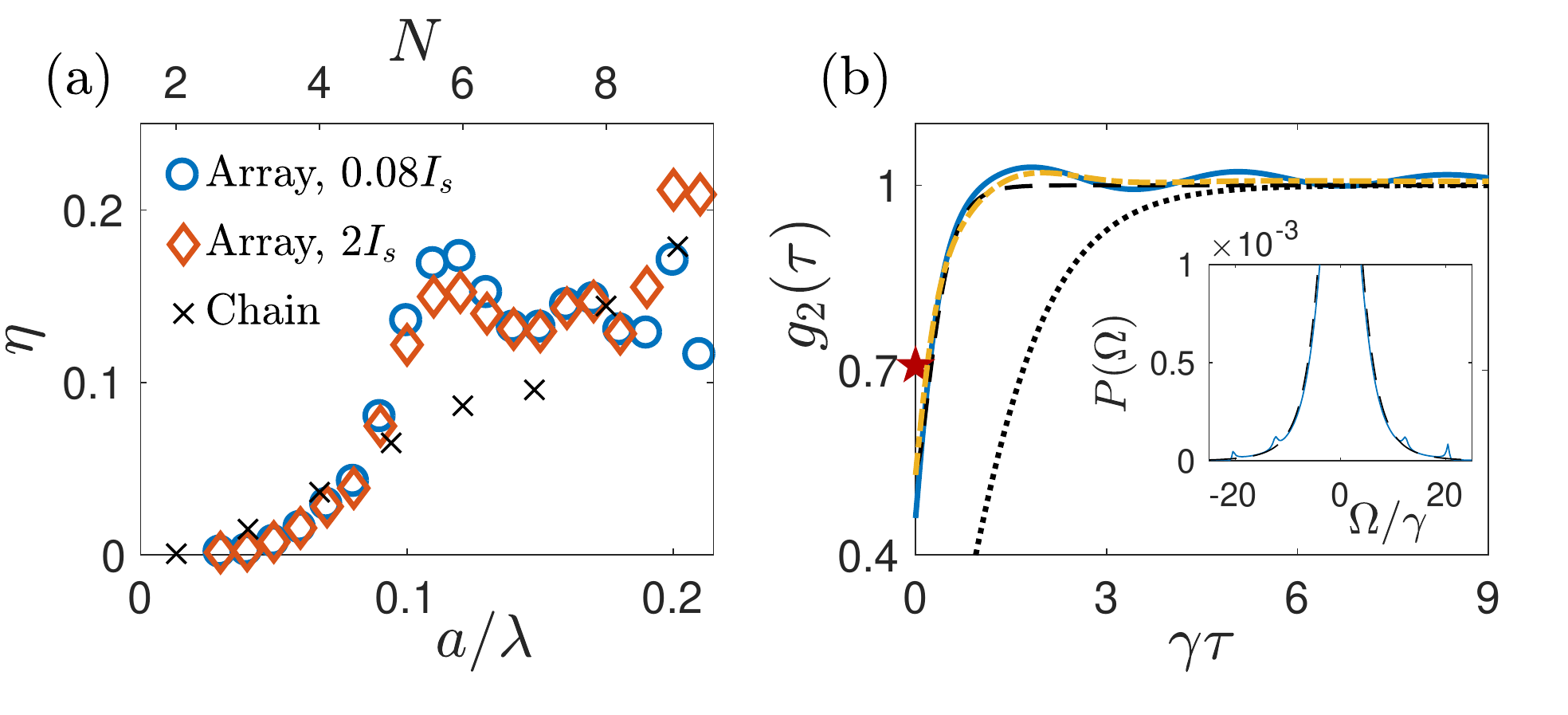}\vspace{-0.5cm}
\caption{\label{validity} Validity of the superatom picture and the effect of position fluctuations of the atoms for a field resonant with the uniform superradiant LLI mode. (a) Relative error $\eta$ of the SAP as a function of lattice spacing (bottom axis) for a $3\times 3$ atom array at $NI_\text{in}=0.08I_s$ (blue circles) and $NI_\text{in}=2I_s$ (red diamonds), and as a function of atom number (top axis, crosses) for a chain at $a=0.15\lambda$, $NI_\text{in}=0.08 I_s$;
(b) position fluctuations of the atoms improve the accuracy of SAP ($a=0.2\lambda$, $2\times 3$ array): fixed atoms (blue solid line),  fluctuating atoms with rms Gaussian density width $0.1a$ at each lattice site (yellow dashed-dotted line), superatom (black dashed line), and single atom (black dotted line). The red star marks the noninteracting, interfering result of $g_2(0)$. Inset: Power spectrum for a $2\times 3$ array ($I_\text{in}=2I_s$, $a=0.08\lambda$) showing a superradiant central peak (SAP result: dashed line) with additional small excitations far off resonance.}
\end{figure}

The suppressed short-delay joint photon detection events in $g_2$ represent dipole blockade that inhibits multiple excitations of the collective atomic state, as illustrated in the excited-state atom number distributions (Fig.~\ref{blockade}).
 Already for a $2\times 3$ array the multiple-excitation probability remains very low at small spacings. While the single-excitation weights are high, e.g., for the lattice spacing $a=0.05\lambda$, the two-excitation weight is $\alt 10^{-5}$ at $NI_\text{in}=2I_s$, but rapidly increases to 0.1 for $a=0.5\lambda$, as the antibunching is reduced and the dipole blockade removed. 
The origin of the blockade can be understood also in the excitation spectrum $P(\Omega)\propto \int d\tau\, e^{i\Omega\tau}\sum_{j\ell}\gamma_{j\ell}\langle\hat{\sigma}_j^+(t+\tau)\hat{\sigma}_\ell^-(t)\rangle$ [inset to Fig.~\ref{validity}(b)] that shows how the second photon excitation is shifted due to the dipole-dipole interactions.

The accuracy and the regimes of validity of the SAP in both planar arrays and chains are analyzed in Fig.~\ref{validity}.
The uniform phase profile of the drive across the atoms most strongly couples to the superradiant, uniform eigenmode, and we show the relative deviations $\eta\equiv\operatorname{max}_{\tau<\tau_0}|g_2(\tau)/b-g_2^{(\upsilon,\kappa^\prime)}(\tau)|$ (calculated until $\tau_0$, such that for all $\tau\alt \tau_0$, $g_2(\tau)<1$; see also Fig.~S2~\cite{supplement}). The SAP describes the behavior of $g_2(\tau)$ very well for $a\alt 0.1\lambda$ and remains qualitatively accurate up to $a\sim 0.2\lambda$ (a 9 atom chain gives similar results). The onset of the plateau around $a\approx 0.12\lambda$, irrespective of light intensity, coincides with LLI eigenmode resonances overlapping with the superradiant mode. For $a \agt 0.2\lambda$, the SAP deviates from $g_2(\tau)$. The deviations as a function of $N$ in Fig.~\ref{validity}(a) show how the accuracy of the SAP decreases gradually in larger systems.

Increasing deviations for large values of $\tau$ for $a\agt 0.2\lambda$ are due to the presence of a persistent oscillation [a weak oscillation is also visible in Fig.~\ref{2d}(a)]. To understand this behavior, we look at the steady-state occupations of the LLI modes for $\langle\sigma_j^-\rangle$, defined as~\cite{Facchinetti16} $L_m\equiv \sum_j |u_m(\mathbf{r}_j)\langle\sigma_j^-\rangle|^2/\sum_{j\ell} |u_\ell(\mathbf{r}_j)\langle\sigma_j^-\rangle|^2$. The presence of the persistent oscillation coincides with a simultaneous nonnegligible occupation of two eigenmodes. One can then qualitatively understand the effect of the two-mode interference from the linear combination
\begin{equation}\label{eq:superatom2}
g_2(\tau)\approx 1+b\left[Cg_2^{(\upsilon_1,\kappa^\prime)}(\tau)+(1-C)g_2^{(\upsilon_2,\kappa_2)}(\tau)-1\right],
\end{equation}
where the increasing contribution from the less radiant mode with increasing lattice spacing leads to deviations from the simple SAP at large $\tau$. Although we consider only chains and arrays, systems with higher symmetry such as rings~\cite{Dong16} or configurations that optimize interactions offer the potential to more effectively target individual superatom resonances and enhance the photon blockade.

For atoms in optical lattices, proposals exist to produce a tight atom confinement~\cite{Wang18}, but generally the atomic positions fluctuate. We can take into account the position fluctuations in the numerics by ensemble-averaging over many stochastic realizations of randomly sampled atom positions in each lattice site~\cite{Jenkins2012a}. We find in Fig.~\ref{validity}(b) that the accuracy of the superatom picture increases due to the fluctuations, as the oscillations resulting from the second eigenmode contribution are washed out. However, increasing position fluctuations eventually also start increasing $g_2(0)$.

\begin{figure}
\includegraphics[trim=0.5cm 10.4cm 2cm 11.5cm,clip=true,width=0.5\textwidth]{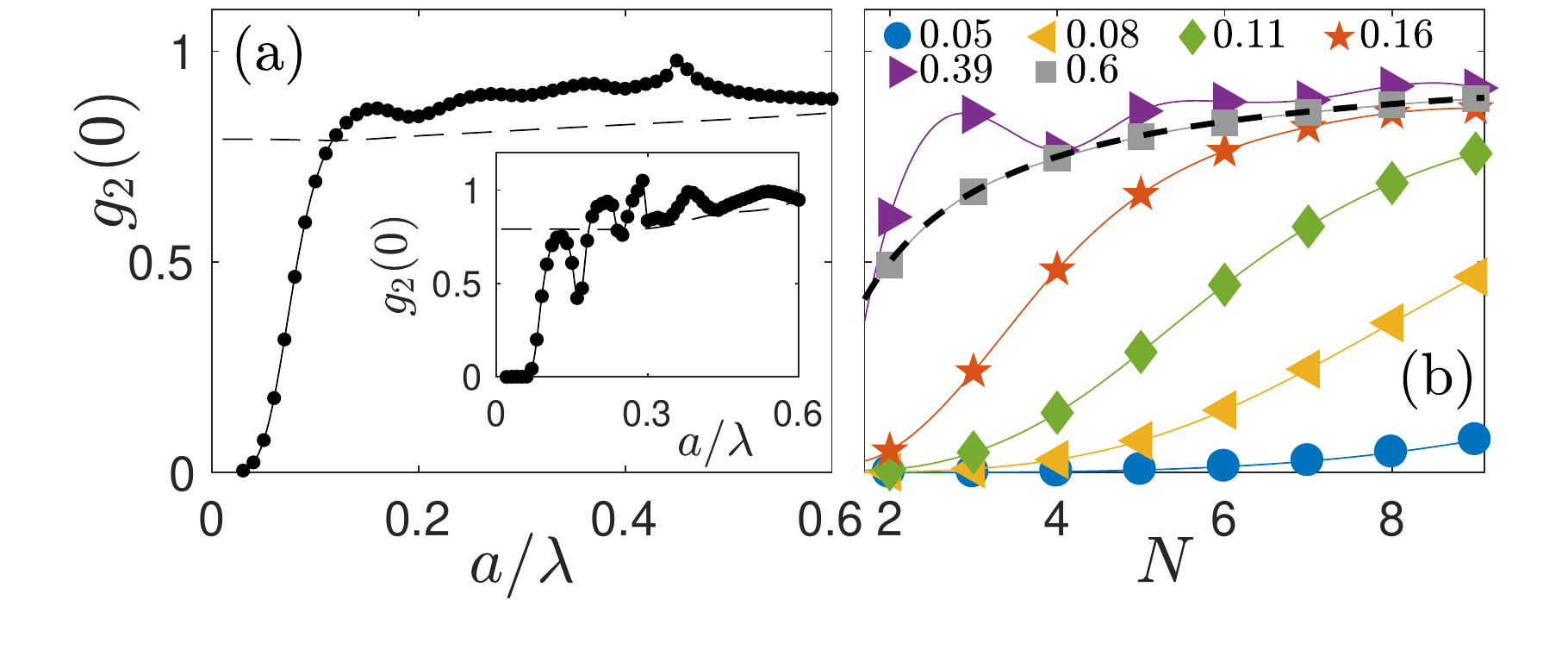}\vspace{-0.6cm}
\caption{\label{fig:g2(0)}Enhanced antibunching due to quantum correlations of light-induced dipole-dipole interactions in a 9-atom chain and  $3\times 3$ array. (a) $g_2(0)$ for a chain as a function of lattice spacing (inset:~array) compared with noninteracting atoms (dashed line);
(b) $g_2(0)$ as a function of atom number for chains with different lattice spacing $a/\lambda$ compared with noninteracting, noninterfering atoms (dashed line). Solid lines are guides for the eye.}
\end{figure}

The normalization of the SAP two-time correlation function at zero delay $g_2(0)$ in Eq.~\eqref{eq:superatom} represents the strength of nonclassical and correlated light emission of the atoms. 
For noninteracting atoms in the absence of multiple scattering, interference effects only slightly modify the result $g_2(0)=1-N^{-1}$. 
Strong light-mediated correlations, however, can substantially shift the value of $g_2(0)$, directly reflected in the antibunching of the emitted photons. In Fig.~\ref{fig:g2(0)}(a) we show $g_2(0)$ as a function of lattice spacing and atom number, with the drive tuned to the uniform LLI eigenmode. We find that light-mediated interactions enhance the nonclassical nature of light for small lattice spacing (up to $a \alt 0.15\lambda$), which coincides with the regime where the SAP shows good accuracy over all values of $\tau$. For chains with large lattice spacing ($a\gtrsim 0.5\lambda$), light-mediated interactions between atoms are no longer sufficient to establish collective correlation effects, and $g_2(0)$ follows the noninteracting, noninterfering scaling $g_2(0)=1-N^{-1}$ [Fig.~\ref{fig:g2(0)}(b)], with small or absent antibunching. In denser arrays, however, we find that nonclassical collective effects persist also as the atom number increases. For example, $g_2(0)\approx 0.08$ for a 9-atom chain with $a=0.05\lambda$.

In Rydberg atoms, dipole blockade inhibits multiple excitations within the \emph{blockade radius} $R$~\cite{Tong04}. Due to the long-range interactions present in our system, $R$ is in general not well defined. However, power-law-fit estimates of the dependence of $g_2(0)$ on the system size can be obtained from Fig. 4(b), resulting in
$R$ of the order of $\lambda$, with a small roughly linear increase in $R$ with decreasing lattice spacing~\footnote{In Fig. 4(b) the $1/r$ interaction is cancelled due to $\hat{\mathbf{d}}\cdot\hat{\mathbf{x}}=1$, but different dipole orientations for which the $1/r$ interaction is nonzero show analogous behavior.}. 
Correlations can be suppressed with a sufficiently broad laser~\cite{takei2016} with increasing contributions from multiple modes [Eq.~\eqref{eq:superatom2}] when the bandwidth notably exceeds $\gamma$.

The time-honoured two-time correlation function \eqref{eq:g2} for joint photon emission events from a single atom reveals nonclassical resonance fluorescence of light~\cite{Carmichael_1976,carmichael1976}. Here we showed that the same functional form also describes emission from strongly coupled arrays of atoms, representing a superatom picture of correlated many-atom resonance fluorescence. 
For a single atom the suppression of joint photon emission events is a direct consequence of the fermionic statistics with $(\hat\sigma^\pm)^2=0$ for the single excitation; after the photon emission the electron is in the ground state and cannot re-emit before being excited again. For a many-atom system, the antibunching with $g_2(0)\simeq 0$ similarly represents the presence of only one excitation, where multiple excitations are inhibited by dipole blockade -- 
reminiscent of fermionic character of multiple photon excitations of atoms in waveguides~\cite{Zhang2018}.

In the superatom picture of  many-atom resonance fluorescence
the strength of the correlations can surprisingly be determined by the underlying LLI collective excitation eigenmodes, even when the atoms are strongly driven by the incident laser. Such an effective collective description is quite different from representing the classical optical response of an atomic ensemble as a superatom in the limit of LLI by coupled collective eigenmodes~\cite{Facchinetti16,Facchinetti18}. Our analysis of the $g_2(\tau)$ correlations illustrates how relatively simple and intuitive representations could possibly more generally be extended to understand strongly correlated many-body phenomena in quantum optics far beyond linearly responding coupled classical dipoles.

 \begin{acknowledgments}
We have become aware of a related parallel theoretical work on the calculation of dipolar blockade in atom chains in Ref.~\cite{Cidrim20}. We acknowledge financial support from the Engineering and Physical Sciences Research Council (Grants Nos.\ EP/S002952/1 and EP/P026133/1) and discussions with L.\ F.\ dos Santos.
\end{acknowledgments}

\end{document}

\title{Supplemental Material to \\``A superatom picture of collective nonclassical light emission and dipole blockade in atom arrays''}
\author{L. A. Williamson}
\affiliation{Department of Physics, Lancaster University, Lancaster LA1 4YB, United Kingdom}
\author{M. O. Borgh}
\affiliation{Faculty of Science, University of East Anglia, Norwich NR4 7TJ, United Kingdom}
\author{J. Ruostekoski}
\affiliation{Department of Physics, Lancaster University, Lancaster LA1 4YB, United Kingdom}
\date{\today}

\maketitle

%% custom supplemental style %%
\renewcommand{\thesection}{S\Roman{section}}
\renewcommand{\thesubsection}{S\Roman{section}.\Alph{subsection}}
\renewcommand{\theequation}{S\arabic{equation}}
\renewcommand{\thefigure}{S\arabic{figure}}
\renewcommand{\bibnumfmt}[1]{[S#1]}
\renewcommand{\citenumfont}[1]{S#1}

%% begin text %%
\subsection{Formalism}

The dynamics in the interaction picture for an array of $N$ two-level atoms driven by a coherent laser field
is described by the many-body quantum master equation (QME) for the reduced density matrix $\rho$~\cite{Lehmberg1970,agarwal1970},
\begin{equation}\label{eq:qme}
  \begin{split}
    \frac{d\rho}{dt}= &-\frac{i}{\hbar}\sum_j[H_j,\rho]+i\sum_{j\ell(\ell\ne j)}\Delta_{j\ell}[\hat{\sigma}_j^+\hat{\sigma}_\ell^-,\rho]\\
    &+\sum_{j\ell}\gamma_{j\ell}\left(2\sigma_j^-\rho\sigma_\ell^+-\sigma_\ell^+\sigma_j^-\rho-\rho\sigma_\ell^+\sigma_j^-\right).
  \end{split}
\end{equation}
Here $\hat{\sigma}_j^+= (\hat{\sigma}_j^-)^\dagger =  |e\rangle_{j}\mbox{}_{j}\langle g|$, $\hat{\sigma}_j^{ee}=\hat{\sigma}_j^+\hat{\sigma}_j^-$ are the atomic raising (lowering) and excited state population operators, with ground $|g\rangle_{j}$ and excited $|e\rangle_{j}$ states of atom $j$. The Hamiltonian operator
\begin{equation}
H_j\equiv -\hbar\delta\hat{\sigma}_j^{ee}-\mathbf{d}\cdot\cbEL^+(\mathbf{r}_j)\hat{\sigma}_j^+-\mathbf{d}^*\cdot\cbEL^-(\mathbf{r}_j)\hat{\sigma}_j^-.
\end{equation}
describes the dynamics of a single atom at position $\mathbf{r}_j$ with the dipole moment $\mathbf{d}\equiv\mathcal{D}\hat{\mathbf{d}}$. Here $\mathcal{D}$ is the reduced dipole matrix element that we assume is real without loss of generality. The atoms are driven by a plane-wave drive with positive frequency component $\cbEL^+(\mathbf{r})=\frac{1}{2}\mathcal{E}_0 e^{i\mathbf{k}\cdot\mathbf{r}}\hat{\mathbf{e}}= [\cbEL^-(\mathbf{r})]^*$.  The drive field frequency $\omega$ is detuned from the single-atom transition frequency $\omega_0$ by $\delta\equiv\omega-\omega_0$. Here the atomic and light fields are slowly varying, such that the rapidly rotating phase factors $e^{\pm i\omega t}$ are removed by moving into an interaction picture and making the rotating wave approximation (by omitting the fast co-rotating terms $\hat{\sigma}_m^- e^{2i\omega t}$, $\hat{\sigma}_m^+ e^{-2i\omega t}$). The single-atom dynamics is thus described by $H_j$ together with the decay terms $\gamma(2\sigma_j^-\rho\sigma_j^+-\sigma_j^+\sigma_j^-\rho-\rho\sigma_j^+\sigma_j^-)$, where $\gamma \equiv \mathcal{D}^2k^3/(6\pi\hbar\epsilon_0)$ is the single atom Wigner-Weisskopf  linewidth.

The scattered light is given as a sum of the scattered light from all the atoms
\begin{align}\label{Escat}
\epsilon_0\hat{\mathbf{E}}_\text{sc}^+(\mathbf{r},t)=\sum_j \mathsf{G}(\mathbf{r}-\mathbf{r}_j)\mathbf{d}\hat{\sigma}_j^-(t)
\end{align}
where the dipole radiation kernel~\cite{Jackson},
\begin{align}
\mathsf{G}(\mathbf{r})\mathbf{d}&=-\frac{\mathbf{d}\delta(\mathbf{r})}{3}+\frac{k^3}{4\pi}\Bigg\{\left(\hat{\mathbf{r}}\times\mathbf{d}\right)\times\hat{\mathbf{r}}\frac{e^{ikr}}{kr}\nonumber\\
&\phantom{==}-\left[3\hat{\mathbf{r}}\left(\hat{\mathbf{r}}\cdot\mathbf{d}\right)-\mathbf{d}\right]\left[\frac{i}{(kr)^2}-\frac{1}{(kr)^3}\right]e^{ikr}\Bigg\},
\end{align}
represents the monochromatic positive frequency component of the scattered light at $\mathbf{r}$ from the dipole $\mathbf{d}$ located at the origin. The interaction terms in Eq.~\eqref{eq:qme} arise from each atom $j$ being driven by the light scattered from all other atoms $\ell\ne j$. These radiative dipole-dipole couplings have coherent $\Delta_{j\ell}$ and dissipative $\gamma_{j\ell}$ contributions given by the real and imaginary parts of
\begin{equation}\label{deltgamdef}
\Delta_{j\ell}+i\gamma_{j\ell}=\frac{1}{\hbar\epsilon_0}\mathbf{d}^*\cdot\mathsf{G}(\mathbf{r}_j-\mathbf{r}_\ell)\mathbf{d}.
\end{equation}
Note that Eq.~\eqref{deltgamdef} gives $\gamma_{jj}=\gamma$. A proper calculation of  $\Delta_{jj}$ would involve evaluation of the Lamb shift, and we assume this is incorporated to the single-atom detuning $\delta$.

The total field at position $\mathbf{r}$ is given as a sum of the incident field $\cbEL^+(\mathbf{r})$ and the scattered light $\hat{\mathbf{E}}_\text{sc}^+(\mathbf{r},t)$. We assume that the incident field has been blocked before detection, for example by a thin
wire as in the dark-ground imaging technique of~\cite{andrews1996}. Hence only the scattered field is detected, with intensity
\begin{equation}
I_\text{sc}(\mathbf{r},t)=2\epsilon_0c\langle\hat{\mathbf{E}}_\text{sc}^-(\mathbf{r},t)\cdot\hat{\mathbf{E}}_\text{sc}^+(\mathbf{r},t)\rangle.
\end{equation}

Integrating the scattered intensity over the detector surface $S$ gives the total count rate, which is the expectation value of the operator
\begin{equation}
\hat{n}(t)=\frac{2\epsilon_0c}{\hbar\omega_0}\int_S dS\,\hat{\mathbf{E}}_\text{sc}^-(\mathbf{r},t)\cdot\hat{\mathbf{E}}_\text{sc}^+(\mathbf{r},t)=\sum_{j,\ell}I_{j\ell}\hat{\sigma}_j^+(t)\hat{\sigma}_\ell^-(t).
\end{equation}
with interference integrals
\begin{equation}
I_{j\ell}\equiv\frac{2c}{\hbar\epsilon_0\omega_0}\int_S dS \left[\mathsf{G}(\mathbf{r}-\mathbf{r}_j)\mathbf{d}\right]^*\mathsf{G}(\mathbf{r}-\mathbf{r}_\ell)\mathbf{d}
\end{equation}

We assume the detector lies in the radiation zone $kr \gg 1$, hence we can expand the dipole radiation kernels to obtain~\cite{carmichael2000} 
\begin{equation}\label{intG}
\begin{split}
I_{j\ell}&=\frac{3\gamma}{4\pi}\int_S d\theta d\phi\,\sin\theta\left(1-|\hat{\mathbf{r}}\cdot\hat{\mathbf{d}}|^2\right)e^{ik\hat{\mathbf{r}}\cdot(\mathbf{r}_j-\mathbf{r}_\ell)}\\
&=2\gamma_{j\ell}.
\end{split}
\end{equation}
Hence we arrive at
\begin{align}\label{nint}
\hat{n}(t)=2\sum_{j\ell}\gamma_{j\ell}\hat{\sigma}_j^+(t)\hat{\sigma}_\ell^-(t)
\end{align}
for the photon-number operator.

\subsection{Single-atom physics}
For a single isolated atom, both the photon detection rate $\left<\hat{n}(t)\right>$ and the second-order correlation function $g_2(\tau)$ can be evaluated analytically to yield~\cite{mollow1969b,Carmichael_1976,carmichael1976,kimble1976}
\begin{equation}\label{eq:singleatom}
\begin{split}
g_2(\tau)&=1-e^{-3\gamma \tau/2}\left(\cosh \kappa \gamma \tau+\frac{3}{2}\frac{\sinh \kappa \gamma \tau}{\kappa}\right)\\
\left<\hat{n}(t)\right>&=\frac{I_\text{in}}{I_\text{in}+I_s}g_2(\tau)
\end{split}
\end{equation}
The parameter $\kappa=\frac{1}{2}\sqrt{1-8I_\text{in}/I_s}$ depends on the ratio of the incident intensity to the single atom saturation intensity, and determines the spectral properties of the atom~\cite{Mollow1969a}. For low incident light intensity $I_\text{in}\ll I_s$, Eqs.~\eqref{eq:singleatom} are dominated by a term proportional to the single decay $e^{-\gamma t}$. Here the single atom linewidth exceeds the single atom Rabi frequency $\gamma\sqrt{2I_\text{in}/I_s}$ and Rabi oscillations are suppressed. Conversely, when $I_\text{in}\gtrsim I_s$, the parameter $\kappa$ is imaginary and hence both the photon scattering rate and $g_2(\tau)$ display decaying Rabi oscillations.

\subsection{Limit of low light intensity}
A consistent low light intensity (LLI) theory of Eq.~\eqref{eq:qme},  can be obtained~\cite{Ruostekoski1997a} from the equations of motions by retaining terms containing at most one of either $\sigma_j^\pm$ or the incident field amplitude. The only remaining equations of motion for the expectation values of atomic operators from Eq.~\eqref{eq:qme} are those for $\langle\sigma_j^\pm\rangle$, which in the LLI are,
\begin{equation}
\frac{d\langle\sigma_j^-\rangle}{dt}=i\delta\langle\sigma_j^-\rangle+i\sum_\ell\mathcal{H}_{j\ell}\langle\sigma_\ell^-\rangle+i\frac{\mathbf{d}\cdot\cbEL^+(\mathbf{r}_j)}{\hbar}
\end{equation}
with $\mathcal{H}_{j\ell}\equiv \Delta_{j\ell}+i\gamma_{j\ell}$ (with $\Delta_{jj}\equiv 0$; recall that $\gamma_{jj}=\gamma$). Hence the atom dynamics evolves linearly in terms of the drive. Here we expand the complex symmetric matrix $\mathcal{H}_{j\ell}$ in a complete basis of eigenstates $u_m$, $m=1,...,N$, which are the LLI collective eigenmodes,
\begin{equation}
\sum_\ell\mathcal{H}_{j\ell}u_m(\mathbf{r}_\ell)=(\zeta_m+i\upsilon_m)u_m(\mathbf{r}_j),
\end{equation}
where the imaginary part, $\upsilon_m$, of the eigenvalue gives the collective linewidth of the eigenmode $u_m$ and the real part the line shift $\zeta_m$ from the single-atom resonance.
Note that the eigenstates $u_m$ are not necessarily orthogonal, however, they do satisfy the biorthogonality condition $\sum_j u_m(\mathbf{r}_j)u_n(\mathbf{r}_j)=\delta_{mn}$ (after appropriate normalization of the $u_m$) apart from possible rare cases when $\sum_j u_m(\mathbf{r}_j)u_m(\mathbf{r}_j)=0$.

Given some steady-state values for the $\langle\sigma_j^-\rangle$, a measure of the occupation of the LLI collective mode $u_m$ is given by~\cite{Facchinetti16} 
\begin{equation}
L_m\equiv \frac{\sum_j |u_m(\mathbf{r}_j)\langle\sigma_j^-\rangle|^2}{\sum_{j\ell} |u_\ell(\mathbf{r}_j)\langle\sigma_j^-\rangle|^2}.
\end{equation}

\subsection{Quantum trajectories}

A direct way to solve the QME~\eqref{eq:qme} is via matrix exponentiation of the density-matrix evolution operator. This is convenient for small atom numbers. For larger systems, however, the size of the density matrix becomes prohibitively large ($\sim \!\!2^{2N}$). A more profitable scaling is to employ the Monte Carlo wavefunction method of quantum trajectories~\cite{dalibard1992,Tian92,Dum92,molmer1993}. The evolution of the density matrix is then represented as the ensemble average of many individual realizations of the evolution of a many-body wavefunction $\psi(t)$, whose size scales as $\sim \!\! 2^N$, under a non-Hermitian Hamiltonian operator
\begin{equation} \label{eq:non-herm}
H_S - \frac{i\hbar}{2}\sum_j\Jop_j^\dagger\Jop_j,
\end{equation}
where $\Jop_j$ are jump operators derived from the dissipative terms of QME and $H_S$ represents Hermitian Hamiltonian evolution. Incoherent evolution is incorporated via stochastic quantum jumps that happen with a probability proportional to the loss of norm of the wavefunction as it evolves under \eqref{eq:non-herm}. One can show that this formalism is exactly equivalent to QME for the operator expectation values~\cite{molmer1993}.

A many-body system supports multiple decay channels and unraveling of the QME into an explicit mixture of pure states subject to stochastic evolution can be done in several different ways, corresponding to different constructions of the jump operators, as long as the full incoherent evolution in Eq.~\eqref{eq:qme} is accounted for. For the driven array of two-level atoms of the QME~\eqref{eq:qme} we follow here the ``source-mode'' quantum trajectory formalism~\cite{clemens2003a,carmichael2000}. 
In the single-excitation limit, these jumps correspond to the emission of photons, while their physical interpretation is more convoluted at sufficiently high light intensities to cause multiple excitations
when the jump operators become formal constructions that do not necessarily correspond to any specific measurement record. They, however, provide a straightforward mapping of Eq.~\eqref{eq:qme} to the evolution of quantum trajectories of state vectors.

To formulate the source-mode jump operators, the matrix $\gamma_{j\ell}$ is diagonalized to find its
eigenvalues $\lambda_j$ and the corresponding eigenvectors $\bv_j =
(b_{1j},\ldots,b_{Nj})^T$.  The jump operators are then defined as
\begin{equation}
  \Jop_j = \sqrt{\lambda_j}\bv_j^T\boldvec{\h\Sigma}, \quad
  \Jop^\dagger_j = \sqrt{\lambda_j}\boldvec{\h\Sigma}^\dagger \bv_j,
\end{equation}
where
\begin{equation}
  \boldvec{\h\Sigma} =
  \mymatrix{\h\sigma_1^-\\\vdots\\\h\sigma_N^-}, \quad
  \boldvec{\h\Sigma}^\dagger =
  \mymatrix{\h\sigma_1^+,&\ldots,&\h\sigma_N^+}.
\end{equation}
Then defining
\begin{equation}
  H_S = \sum_j H_j+\hbar\sum_{j\ell(\ell\ne j)}\Delta_{j\ell}\hat{\sigma}_j^+\hat{\sigma}_\ell^-,
\end{equation}
the problem has been cast in the form of quantum trajectories and the corresponding non-Hermitian Hamiltonian for the wavefunction evolution follows from Eq.~\eqref{eq:non-herm}.
The quantum trajectory evolution can then be evaluated as described in, e.g., 
Ref.~\cite{ruostekoski1998}.
Thanks to the source-mode unraveling, the dissipative component of Eq.~\eqref{eq:qme} is now diagonal in the jump operators $\hat{J}_j$, which is computationally expedient. Further, the stochastic wavefunction evolution requires exponentiating a matrix of size $2^{2N}$, as opposed to the $2^{4N}$ matrix governing the density matrix evolution, providing a significant numerical advantage as the system size increases beyond a few atoms.

\begin{figure}
\includegraphics[trim=0cm 10cm 0cm 10cm,clip=true,width=0.5\textwidth]{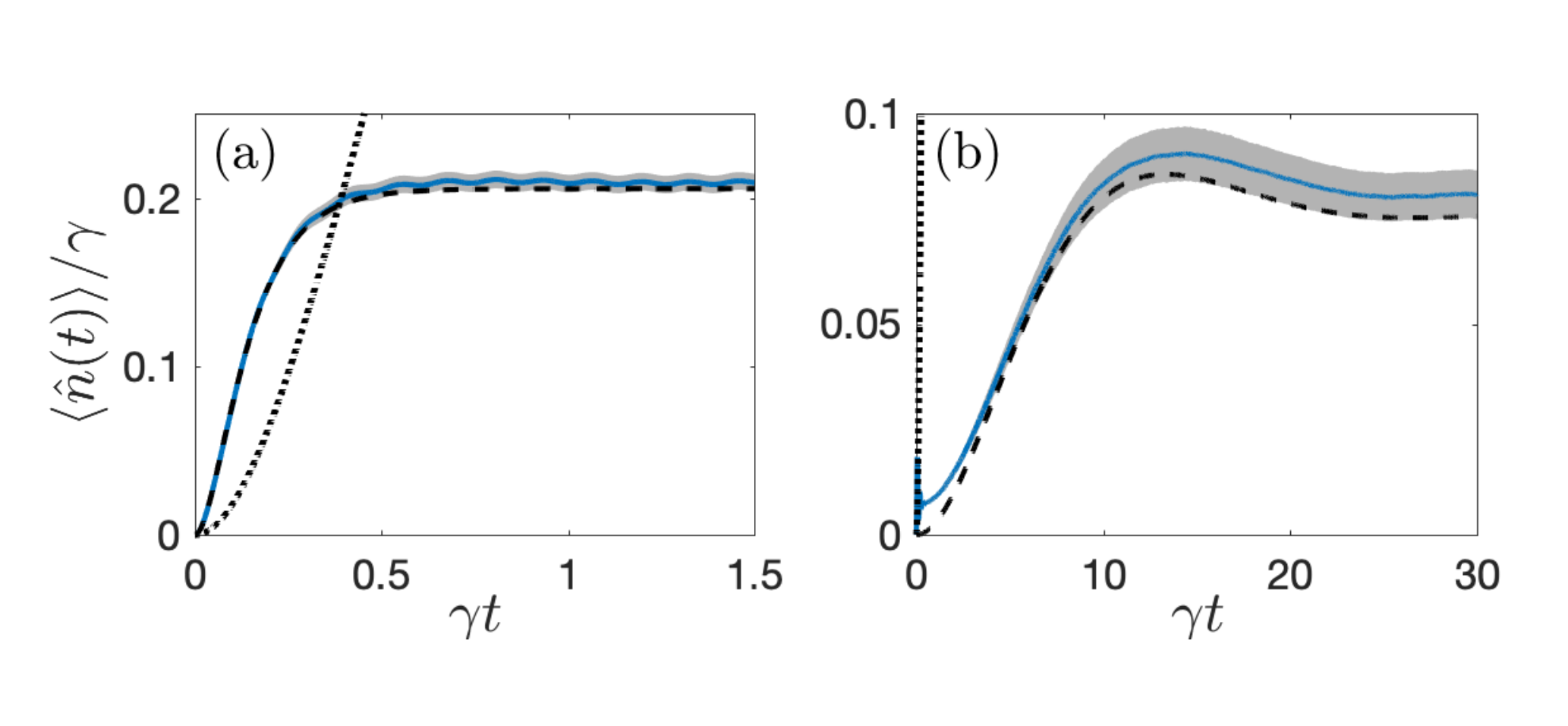}\vspace{-0.6cm}
\caption{\label{posFluc}Transient dynamics of the photon detection rate for a $3\times 4$ atom array with a drive field resonant with (a) the uniform superradiant ($\upsilon\approx 9.3\gamma$, $\mathcal{I}\approx 0.98N$) and (b) a subradiant ($\upsilon\approx 0.11\gamma$, $\mathcal{I}\approx 0.015N$) LLI collective eigenmode, with $NI_\text{in}=2I_s$, $a=0.1\lambda$. The full quantum solution (blue solid line) agrees very well with the superatom (black dashed line); black dotted line shows the single isolated atom solution. The gray shading gives the standard error from $\sim 10^4$ quantum trajectories. Interestingly, examining just the incoherent contribution to the scattering rates in (b) gives even better agreement between the SAM and full quantum solution.}
\end{figure}

\begin{figure}
\includegraphics[trim=0cm 7cm 0cm 10cm,clip=true,width=0.5\textwidth]{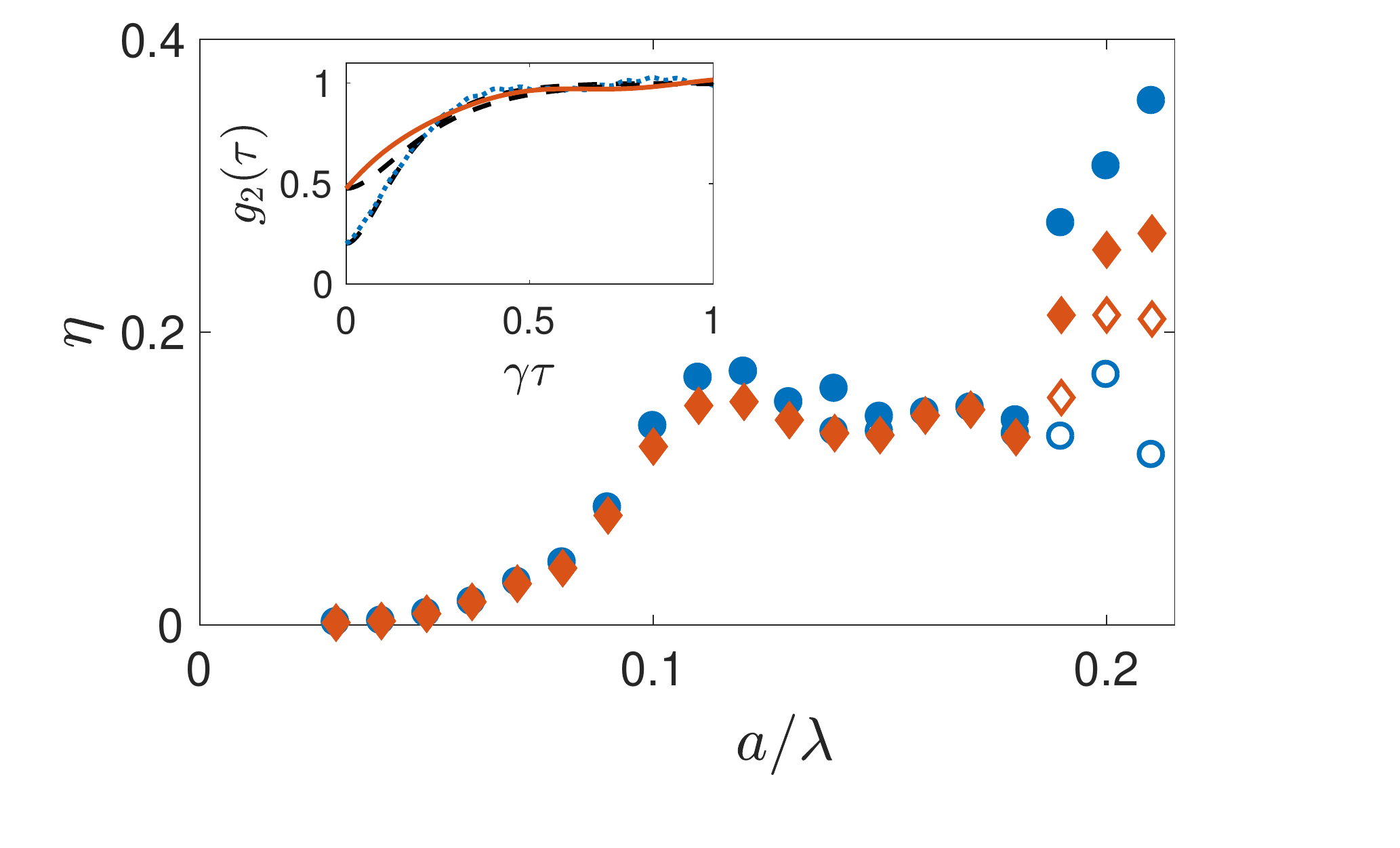}\vspace{-0.7cm}
\caption{Relative error $\eta$ of the superatom picture as a function of lattice spacing for a field resonant with the uniform superradiant LLI mode for a $3\times 3$ atom at $NI_\text{in}=0.08I_s$ (blue circles) and $NI_\text{in}=2I_s$ (red diamonds). Unfilled markers show results for $\eta\equiv\operatorname{max}|g_2(\tau)/[1-g_2(0)]-g_2^{(\upsilon,\kappa^\prime)}(\tau)|_{\tau<\tau_0}$, where the deviation is calculated until $\tau_0$, such that for all $\tau\alt \tau_0$, $g_2(\tau)<0$, while filled markers for $\eta\equiv\operatorname{max}|g_2(\tau)/[1-g_2(0)]-g_2^{(\upsilon,\kappa^\prime)}(\tau)|_\text{all $\tau$}$. The two deviate at $a\sim 0.2\lambda$ due to a persistent oscillation arising from a second mode at larger $\tau$. Inset: Example $g_2(\tau)$ for $a=0.08\lambda$ (blue dotted curve) and $a=0.17\lambda$ (red solid curve) compared to the SAP results (black dashed curves), with the larger lattice spacing showing an oscillation.}
\end{figure}